\documentclass[pra,aps,twocolumn]{revtex4}
\usepackage{epsfig}
\usepackage{amsmath}
\usepackage[dvipsnames]{color}

\begin{document}

%\setstretch{2} 

\title{Optical sum-frequency generation in whispering gallery mode resonators}

\author{Dmitry V. Strekalov}
\affiliation{
Jet Propulsion Laboratory, California Institute of
Technology, 4800 Oak Grove Drive, Pasadena, California 91109-8099}
\author{Abijith S. Kowligy}
\author{Yu-Ping Huang}
\author{Prem Kumar}
\affiliation{Center for Photonic Communication and Computing, EECS Department, Northwestern University,
2145 Sheridan Road, Evanston, Illinois 60208-3118, USA}

\date{\today}

\begin{abstract}

We demonstrate sum-frequency generation in a nonlinear whispering gallery mode resonator between a telecom wavelength and the Rb D2 line, achieved through natural phase matching. Due to the strong optical field confinement and ultra high $Q$ of the cavity, we achieve a 1000-fold enhancement in the conversion efficiency compared to existing waveguide-based devices. The experimental data are in agreement with the nonlinear dynamics and phase matching theory in the spherical geometry employed.  The experimental and theoretical results point to a new platform to manipulate the color and quantum states of light waves toward applications such as atomic memory based quantum networking and logic operations with optical signals.

\end{abstract}


\maketitle

Strong optical nonlinearities have been the foundation of many applications in classical and quantum optics. Recently, the burgeoning field of high-$Q$ nonlinear micro- and nano-cavities \cite{Vahala03} has emerged as a new chip-scalable platform for photonic information processing, which requires very low ($<$1 fJ) energies \cite{Nozaki10, Huang10, Huang12a} and can be nearly lossless. In addition, by utilizing the quantum Zeno effect, interaction-free operations can be implemented, which eliminates the otherwise inevitable energy dissipation and background scattering processes. A pursuit of low-light level optical interactions, hence, can simultaneously address fundamental and practical problems faced by both classical and quantum information processing \cite{Miller}. In fact, by analyzing a $\chi^{(2)}$-nonlinear Lithium Niobate microresonator, it has been shown that strong, noise-free interaction can be realized among single photons, thereby uncovering pathways to unprecedented applications such as optical transistors and deterministic quantum logic gates \cite{Sun13}. Such a realization has an inherent advantage over resonant optical interactions with matter systems due to its compact experimental setup and a room-temperature operation. 

All of these proposals place exacting criteria on the resonators, requiring a high quality factor, small mode volume, and good overlap between the interacting modes. While a small mode volume is available in photonic-crystal microcavities,  multiply-resonant, high-$Q$ cavities are difficult to fabricate \cite{Majumdar13}. Here, we present for the first time, naturally phase-matched sum-frequency generation (SFG) in a triply-resonant high-$Q$ Lithium Niobate microresonator with strongly non-degenerate added frequencies. %This demonstration serves as a significant step towards achieving these goals simultaneously. 
Optically nonlinear resonators thus far have been successfully used for either single frequency multiplication (e.g., doubling \cite{ilchenko04SH,furst10,Kuo11}, tripling \cite{carmon07,Sasagawa09}, and quadrupling \cite{moore11}), or parametric down conversion \cite{Savchenkov07,furst10pdc,beckmann12,Werner12}. In contrast, in our experiment we demonstrate SFG between a 1560 nm pump and a 780 nm signal. Such cross band coupling opens avenues to several narrow-band frequency conversion applications that have been hitherto challenging. Indeed, SFG can be employed for efficient room-temperature detection of far-infrared, and even sub-THz, photons \cite{StrekalovTHz,Ma12}.  Since SFG does not disturb the quantum state \cite{QFC}, it also can lead to efficient manipulation of the color and shape of single-photon signals \cite{Ray12} for interfacing optical flying qubits with narrowband atomic quantum memories \cite{Shahriar12}. Furthermore, the narrow-band resonance lines in such devices can greatly suppress incoupling of Raman noise, and potentially lead to new optical tools for mode discrimination and reshaping of narrowband quantum signals \cite{Brecht}. Our experiment is an important first step towards all of these applications in both the classical and quantum domains. 

% \textbf{Experimental setup!!!}
\begin{figure}[htb]
%\vspace*{-0.1in}
\centerline{ %\hspace*{-0.5in}
\includegraphics[width=9cm]{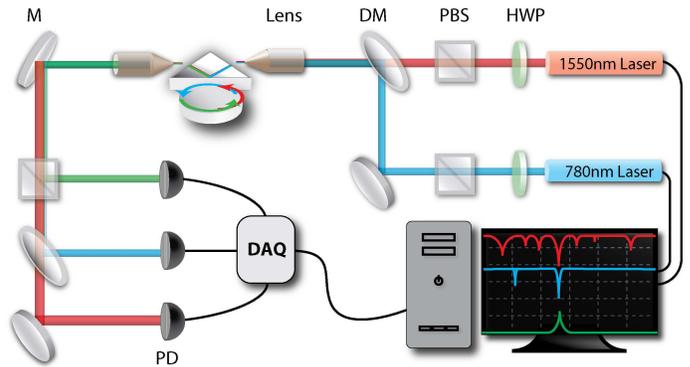}
}
%\vspace*{2.5in}
\caption[]{\label{fig:setup}Experimental setup. PD: Photodetector, M: Mirror; DM: Dichroic Mirror; PBS: Polarization Beam Splitter; HWP: Half-Wave Plate; DAQ: Data Acquisition Unit. 10X Objective Lenses were used to focus the lasers onto the prism-resonator interface and to collect the output light. }
\end{figure}

Our experiment is illustrated in Fig.~\ref{fig:setup}. We observed sum-frequency generation in a MgO-doped Lithium Niobate z-cut microdisk (R $\approx 0.6$ mm) evanescently coupled to a diamond prism. Diamond-polishing was used to obtain absorption-limited $Q \ge 2\times 10^7$ and $Q \ge 4\times 10^7$ for the signal and pump waves, respectively. % Uncoated prism surfaces resulted in an approximately $3$dB transmission loss for our setup. 
A 780nm narrow (below $300$ kHz) linewidth tunable-diode laser was used as the signal and a DFB laser provided 1560nm pump. Two input waves are ordinarily polarized to achieve the Type-I phase-matching whereas the upconverted wave at 520nm is detected in the extraordinarily polarization. The two input waves were combined on a dichroic mirror and focused onto the prism-resonator interface by an objective lens. A lateral offset of the beams before the lens allowed for optimizing of the pump and probe in-coupling angles individually. A similar lens was used to collect the output light, with a dichroic mirror separating the pump and the signal while a polarization beam splitter separating the sum-frequency wave. All three optical powers were measured by photodetectors, whose signals were fed into a data acquisition unit. As the laser frequencies were continually swept at 50Hz across several linewidths, the signal and pump WGMs were tracked by a software program continuously adjusting the lasers central wavelength to follow their respective WGMs. The program ensured that the pump and signal WGMs were in the center of the sweeps and that they were pumped simultaneously. In addition, the top of the resonator was coated with silver paste and temperature-controlled to allow for electro-optic and thermal tuning for the SFG phase matching. 

% Experimental results 
Having the phase matching achieved, we measured the sum-frequency output power for various input pump and signal powers. Since the temperature stabilization of the resonator at the level of the phase matching temperature width (approximately 7 mK) was deemed difficult and time consuming, we carried out these measurements in transient by slowly varying the electro-optic bias voltage to record the peak SFG efficiencies. Each data point represents the average of three consecutive measurements.

The signal and pump waves were critically coupled and over-coupled, respectively. The longer wavelength pump wave was coupled stronger than the signal due to the nature of the evanescent coupling. Also, due to the spatial-mode mismatch between the input Gaussian beam and the WGM profile, in this measurement we achieved a critically-coupled contrast of 48\%. We took this into account by only utilizing the in-coupled powers for our theoretical analysis. 

\begin{figure}[htb]
%\vspace*{-0.1in}
\centerline{ %\hspace*{-0.5in}
\includegraphics[width=9cm]{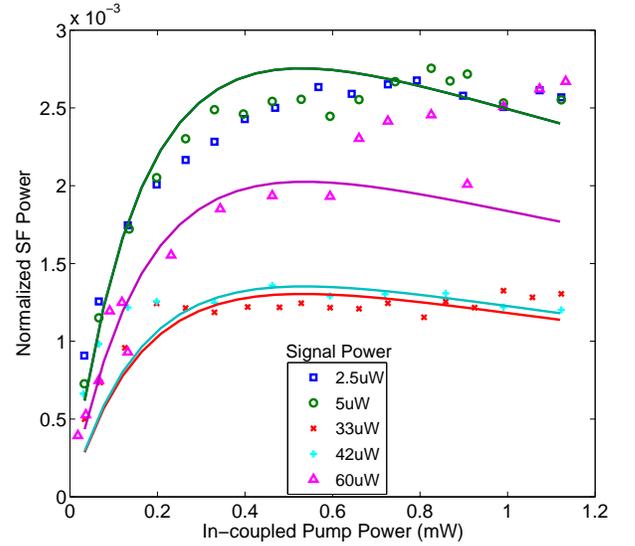}
}
%\vspace*{-0.2in}
\caption[]{\label{fig:effcy}The out-coupled SF emission is measured and normalized to the various input signal powers. Symbols represent the experimentally measured data, and the solid lines are theoretical fits. Note that the theory predictions for 2.5$\mu W$ and 5$\mu W$ are identical because the nonlinear loss is negligible at such powers.}
\end{figure}

We observed efficient sum-frequency generation with a maximum in-coupled pump power of only 1.22 mW. In Fig.~\ref{fig:effcy} we plot the \emph{out-coupled SFG} efficiency. %Since we cannot measure the SFG out-coupling rate, we could not directly evaluate the internal conversion efficiency. However this rate can be inferred from the input-output power curves, as will be shown below.  
As we varied the signal powers, saturation of the peak conversion efficiency was observed in all cases with sub-milliwatt incoupled powers, instead of a cyclic behavior that is observed in the traveling-wave configuration \cite{Boyd}. Similar saturation has been observed in frequency-doubling WGM experiments \cite{ilchenko04SH,furst10}. At higher pump powers, an additional nonlinear loss for the signal wave is created due to its upconversion, leading to the reduction of its internal $Q$-factor and the coupling contrast. As a result, a smaller portion of the signal wave enters the resonator and the SFG efficiency is reduced. This behavior is a manifestation of the ``coherent" quantum Zeno effect for the signal wave, where the ``potential" for the upconversion decouples the signal field from the cavity \cite{Huang12a}. 

%%%%%%%%  Dynamics - theory
A theoretical description of the SFG in WGMRs is warranted because manifestly different behavior is observed at high pump powers compared to a traditional traveling-wave geometry. Neglecting Rayleigh backscattering and assuming linearly polarized fields, we find the scalar equations of motion in the cavity are:
\begin{eqnarray}
\frac{\partial c_{p}}{\partial t} &= &-\kappa_{p} c_{p} + i\sqrt{\frac{\omega_p}{Q^{c}_{p}}}a_{p} + i\Omega c_{s}^{*}c_{f}\\
\frac{\partial c_{s}}{\partial t} &= &-\kappa_{s} c_{s} + i\sqrt{\frac{\omega_s}{Q^{c}_{s}}}a_{s} + i\Omega c_{f}c^{*}_{p}\\
\frac{\partial c_{f}}{\partial t} &= &-\kappa_{f} c_{f} + i\Omega^{*} c_{s}c_{p}
\end{eqnarray}
where $Q^{c}_{\mu}$ and $Q^{i}_{\mu}$ denote the coupling and intrinsic $Q$-factors, $\kappa_{\mu}=(\frac{\omega_{\mu}}{2Q^{i}_{\mu}} + \frac{\omega_{\mu}}{2Q^{c}_{\mu}} - i\Delta_{\mu})$ for $\mu=s,p,f$ indicating respectively signal, pump and the sum-frequency, and $\Omega=\frac{\epsilon_{0}}{\hbar}d_{31}\int dV E_{f}E^{*}_{p}E^{*}_{s}$ is the internal conversion efficiency of the SFG process.  The input field operators, $a_{\mu}$, are related to the output fields by $b_{\mu} = \sqrt{T}a_{\mu} + i\sqrt{\frac{\omega}{Q^{c}_{\mu}}}c_{\mu}$. Using quasi-static analysis we solve Eqns. (1)-(3) for the out-coupled sum-frequency field, $|b_{f}|^{2} = |i \sqrt{\frac{\omega_{f}}{Q_{f}^{c}}} \frac{\Omega}{ \kappa_{f} }c_{s} c_{p}|^{2}$. 

Before we provide the solution, a few salient points are to be noted. In this formulation, $\Omega$ is the \emph{internal} conversion efficiency of the SFG process, whereas only the SF \emph{out-coupled} power is measured experimentally. Since $Q_{\mu}^{c}$ is determined by the distance from the prism to the resonator $d$, we need only to fit to $\Omega$ and the intrisic $Q$-factor for the SF \cite{suppl}. Moreover, these equations are difficult to solve in general for $c_s$, $c_p$, and $c_f$ analytically. To guide us, however, we make the undepleted pump approximation and then use numerical methods to acquire the generic solution. This approximation yields:

\begin{eqnarray}
c_{p} &= &\frac{i\sqrt{\frac{\omega_p}{Q^{c}_{p}}}a_{p}}{\kappa_{p}}, c_{s} = \frac{ i\sqrt{\frac{\omega_s}{Q^{c}_{s}}}a_{s} }{\kappa_{s} + \frac{|\Omega|^2 \omega_{p} |a_{p}|^2}{\kappa_{f}Q_{p}^{c} |\kappa_{p}|^2}}, \\
|b_{f}|^2 &= &\frac{\omega_s \omega_p \omega_f}{Q_{s}^{c}Q_{p}^{c}Q_{f}^{c}} \frac{|\Omega|^2/|\kappa_{f}\kappa_{p}|^2}{|\kappa_s + \frac{|\Omega|^2 \omega_p |a_{p}|^2}{\kappa_f Q_{p}^{c} |\kappa_p|^2}|^2}|a_{p}|^2|a_{s}|^2.\label{bf}
\end{eqnarray}
Note that the expressions for $c_p$ and $c_s$ are asymmetric due to the nature of the undepleted pump approximation. These solutions indicate that we do not observe an oscillatory behavior in the frequency-conversion dynamics. For low pump and signal energies, the SFG output behaves linearly, $|b_f|^2 \propto |a_p|^2 |a_s|^2$ (see Fig.~\ref{fig:effcy}), whereas at higher energies, the dynamics are different, i.e.,  the upconversion and subsequent down conversion processes are asymmetric in this geometry \cite{Huang12a}. 

%Eq.~(\ref{bf}) can be used to empirically infer $\Omega$ for our resonator.
%The theory we have developed is able to faithfully reproduce the significant features of this experiment, including the saturation-type behavior and the reduction of the sum-frequency emission at higher input pump power. 
Using the in-coupled pump and signal powers, and measured $Q_\mu^{i}$,$Q_\mu^{c}$ for $\mu=s,p$, we can provide a theoretical fit for the measured data in Fig.~\ref{fig:effcy} with the fitting parameters $\Omega$ and $Q_{f}^{i}$. Fitting our efficiency measurements at 2.5$\mu$W signal inputs (where the undepleted pump approximation is valid) with Eq.~(\ref{bf}), we can estimate $\Omega$, which does not vary as the same WGM triplet is studied throughout the experiment. Throughout this procedure, we assumed that the SFG peak occurred when the three waves were exactly on resonance, i.e., $\Delta_s = \Delta_p = \Delta_f = 0$.

We then solved for Eqs. (1)-(3) numerically, which provided the efficiency curves for data where the undepleted pump approximation is invalid.  The total $Q$-factors for the pump and signal waves were not constant through the experiment, which we account for by using $Q_{f}^{i}$ as a fitting parameter in Fig.~\ref{fig:effcy}. We attribute this variation in part to the unaccounted photorefractive effects in the resonator, caused by the SF. These effects were most evident for the highest signal power used, $P_s = 60\mu W$. Indeed, this particular data set in Fig.~\ref{fig:effcy} also presents the largest data scatter and the worst agreement with theory. 

In spite of these theoretically unaccounted background processes, by using just the $\Omega$-parameter, which determines the shape of the efficiency curves, we are able to accurately model the signal and pump transmission spectra as well as the SF emission spectrum \cite{suppl}. Moreover, we were able to calculate  \cite{suppl} the fundamental channels modes overlap to give $\Omega_{theor} = 253$ kHz, whereas the empirical value gives us $\Omega_{expt} = 5$ kHz. In calculating this latter value, we used $Q_{f}^{i} = 3.25\times 10^7$, the intrinsic sum-frequency $Q$-factor and that it was strongly undercoupled to the resonator, $\frac{Q_f^c}{Q_f^i}=164$, see \cite{suppl}. %We believe these latter two values, obtained through our fit, are reasonable because the $Q^{i}$ at a close wavelength and in the same material was measured \cite{furst10} to be $4\times 10^7$, and the undercoupling factor order-of-magnitude estimate \cite{suppl} is $\frac{Q_f^c}{Q_f^i}\approx 130$.  The discrepancy between $\Omega_{theor}$ and $\Omega_{expt}$ suggests that the WGM triplet in the experiment may not have been the fundamental one, as we discuss below. 

% Phase matching theory

Not every pair of signal and pump WGMs can generate sum-frequency.
SFG in a triple-resonant system requires the phase matching between these modes, which can be viewed as conservation of the integrals of motion determined by the system's symmetry. 

Usually WGMs have nearly perfect spherical symmetry, so their eigenfunctions inside the resonator can be well approximated by 
\begin{equation}
\Psi_{Lmq}(r,\theta,\varphi)= Y_{L}^m(\theta,\varphi)j_L(k_qr),
\end{equation}
where $r,\theta,\varphi$ are spherical coordinates, $L,m,q$ are azimuthal, polar and radial mode numbers, respectively, $Y_{L}^m(\theta,\varphi)$ is a spherical harmonic and $j_L(k_qr)$ is a spherical Bessel function. The radial wave number $k_q$ is determined from the boundary conditions. 

In \cite{suppl} we show that many combinations of WGM triplets (called \emph{channels} in the following) may lead to SFG albeit with different conversion efficiencies, whereas most efficient channel is the one that couples the fundamental modes, i.e. such that $q_p=q_s=q_f=1$, $L_p-m_p=L_s-m_s=L_f-m_f=0$. 

Achieving a triple resonance for a selected channel requires tuning of each WGM's frequency such that the energy conservation $\omega_p(L_p,m_p,q_p)+\omega_s(L_s,m_s,q_s) = \omega_{f}(L_f,m_s+m_p,q_f)$ is fulfilled to better than a WGM linewidth. In lithium niobate resonators this can be achieved for ordinary pump and signal and extraordinary sum-frequency WGMs due to different temperature dependencies of the ordinary and extraordinary refraction indices. We find the phase matching temperatures \cite{suppl} by iteratively solving the energy conservation condition using the WGM dispersion equation \cite{Gorodetsky} and temperature-dependent Sellmeier equations \cite{Schlarb94}. Let us point out that for sub-mm resonators the resulting temperatures may significantly (by tens of degrees) differ from the bulk phase matching temperature, due to the geometrical, or waveguide, part of the WGM dispersion.

\begin{figure}[htb]
%\vspace*{0.1in}
\centerline{ \hspace*{-0.5in}
\includegraphics[width=9cm]{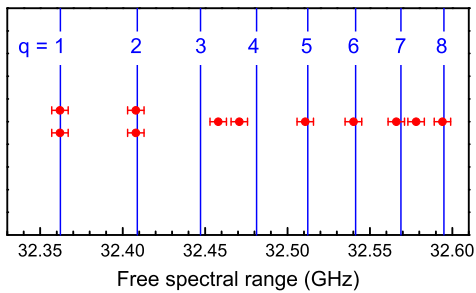}
}
%\vspace*{-0.4in}
\caption[]{\label{fig:FSR}Theoretical FSR values for $q = 1...8$ WGMs at 1560 nm (vertical lines), and the measurement results.}
\end{figure}

In the experiment, identifying a WGM's $q$ may present a considerable difficulty. Fortunately, the WGM's free spectral range (FSR) depends on its $q$ much stronger than on $L$ and $m$, because $q$ affects the effective length of the resonator.  Therefore a WGM's $q$ can be inferred from the FSR measurements. We carried out the FSR measurements for the pump laser by frequency-modulation technique \cite{Li12FSR}. The results of such measurements carried out with 11 best-coupled modes within one FSR are shown in Fig.~\ref{fig:FSR}. The theoretical FSR values shown in Fig.~\ref{fig:FSR} were derived from the WGM dispersion equation \cite{Gorodetsky}. We fit the theory value for $q = 1$ to the smallest measured FSR of 32.362 GHz by varying the resonator radius from the initially measured $0.65\pm 0.01$ mm to $0.6505$ mm. No fitting was done to match other $q$ values. 

% from Nov 12 2012
%

%%%%%%%%%   HERE - want to say that we stepped the lases by N FSRs
The theoretical FSR value for $q=1$ WGMs at the signal wavelength is 31.049 GHz. We found a high-contrast mode with a very close FSR value of $31.04 \pm 0.004$ GHz \cite{rogue_modes}. Coupling the pump and signal lasers to these WGMs we slowly varied the resonator temperature while monitoring the SFG signal, and acquired the data for Fig.~\ref{fig:effcy}. 

% from Numeric simulations\Tuning
\begin{figure}[t]
%\vspace*{-0.1in}
\centerline{ %\hspace*{-0.5in}
\includegraphics[width=10.5cm]{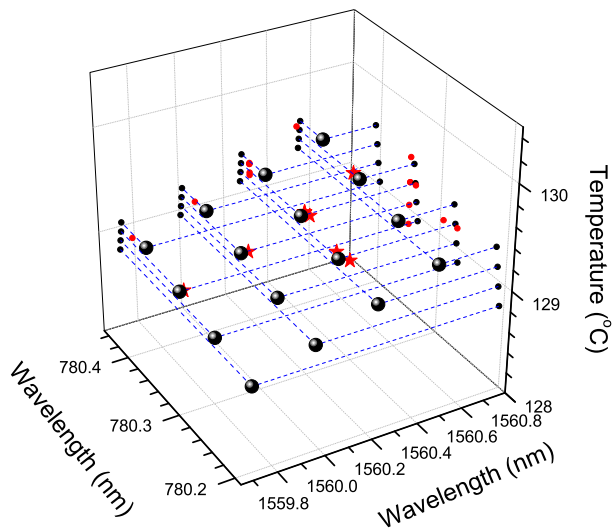}
}
%\vspace*{-0.2in}
\caption[]{\label{fig:tuning}Numeric simulation result for the fundamental SFG channel (black dots) and experimental observations (red stars). Projections emphasize a good agreement between the theory and experiment. }
\end{figure}

We also found the neighboring SFG channels by tuning the pump and signal lasers across an integer number of FSRs. The wavelengths and phase matching temperatures for these channels are shown in Fig.~\ref{fig:tuning} together with the numeric simulation for the fundamental channel. A good agreement was achieved by using the MgO concentration as a fitting parameter in the simulation. This parameter affects the phase matching temperature by entering Lithium Niobate dispersion \cite{Schlarb94}. Unfortunately its value is not precisely known for our congruent wafer, except that it should slightly exceed the threshold value of approximately 5\%. The fitting yielded a very plausible value of 5.63\%. It should be pointed out that no other efficient SFG channels can fit the observations with any reasonable MgO concentration. However, many less efficient equatorial channels exist for the same temperatures \cite{suppl}. The lower observed $\Omega$ suggests that the WGM triplet may not have been fundamental. %This suggestion is made plausible by the uncertainty in the 520nm wave's WGM numbers.  

%%%%%%%%%%%%%%%%%%%%%%%%%%%%%%%%%%%%%%%%%%%%%%%%%%%%%%

To summarize, we have demonstrated triple-resonant SFG in a WGM resonator. We have extended the theoretical analysis for finding the phase-matched WGMs inside the resonator and understood the nonlinear dynamics of frequency conversion in the strong pump-signal coupling regime. The efficiency of this process in the resonator is much higher than in a traveling-wave geometry, requiring sub-milliwatt powers for saturation. We expect to find applications of this interaction in the fields of spectroscopy, optical communications and data processing, both at classical and at quantum levels. Other conceivable applications are for fundamental tests of quantum theory, e.g. in cavity optomechanics and quantum non-demolition measurements. % of the photon can be studied \cite{qnd} and such phonon-photon coupling can also be useful for studying the fundamental mechanisms of quantum decoherence \cite{optodecohr}.
% Furthermore, strong optical field coupling mediated an off-resonant process can also lead to robust all-optical logic gates for both classical and quantum optical communications. 

This work was supported by the DARPA Zeno-based Opto-Electronics program (Grant No. W31P4Q-09-1-0014). It was partly carried out at the Jet Propulsion Laboratory, California Institute of Technology under a contract with the National Aeronautics and Space Administration. We thank J. U. F\"urst and T. Beckmann for useful discussions.


\begin{thebibliography}{99}
\bibitem{rogue_modes}
The majority of the signal modes had smaller FSRs and did not follow the trend of Fig.~\ref{fig:FSR}. This contradicts both the assumption that the selected probe mode has $q=1$ and the resonator radius measurement. Localized photorefractive damage induced by the 780 nm light is one possible explanation of this discrepancy.
\end{thebibliography}
\end{document}

%\setstretch{2} 

\title{Optical sum-frequency generation in whispering gallery mode resonators - Supplementary Material}

\author{Dmitry V. Strekalov}
\affiliation{
Jet Propulsion Laboratory, California Institute of
Technology, 4800 Oak Grove Drive, Pasadena, California 91109-8099}
\author{Abijith S. Kowligy}
\author{Yu-Ping Huang}
\author{Prem Kumar}
\affiliation{Center for Photonic Communication and Computing, EECS Department, Northwestern University,
2145 Sheridan Road, Evanston, Illinois 60208-3118, USA}

\date{\today}

\maketitle

\section{Acquiring $\Omega$}

\begin{figure}[htb]
%\vspace*{-0.1in}
\centerline{ %\hspace*{-0.5in}
\includegraphics[width=9cm]{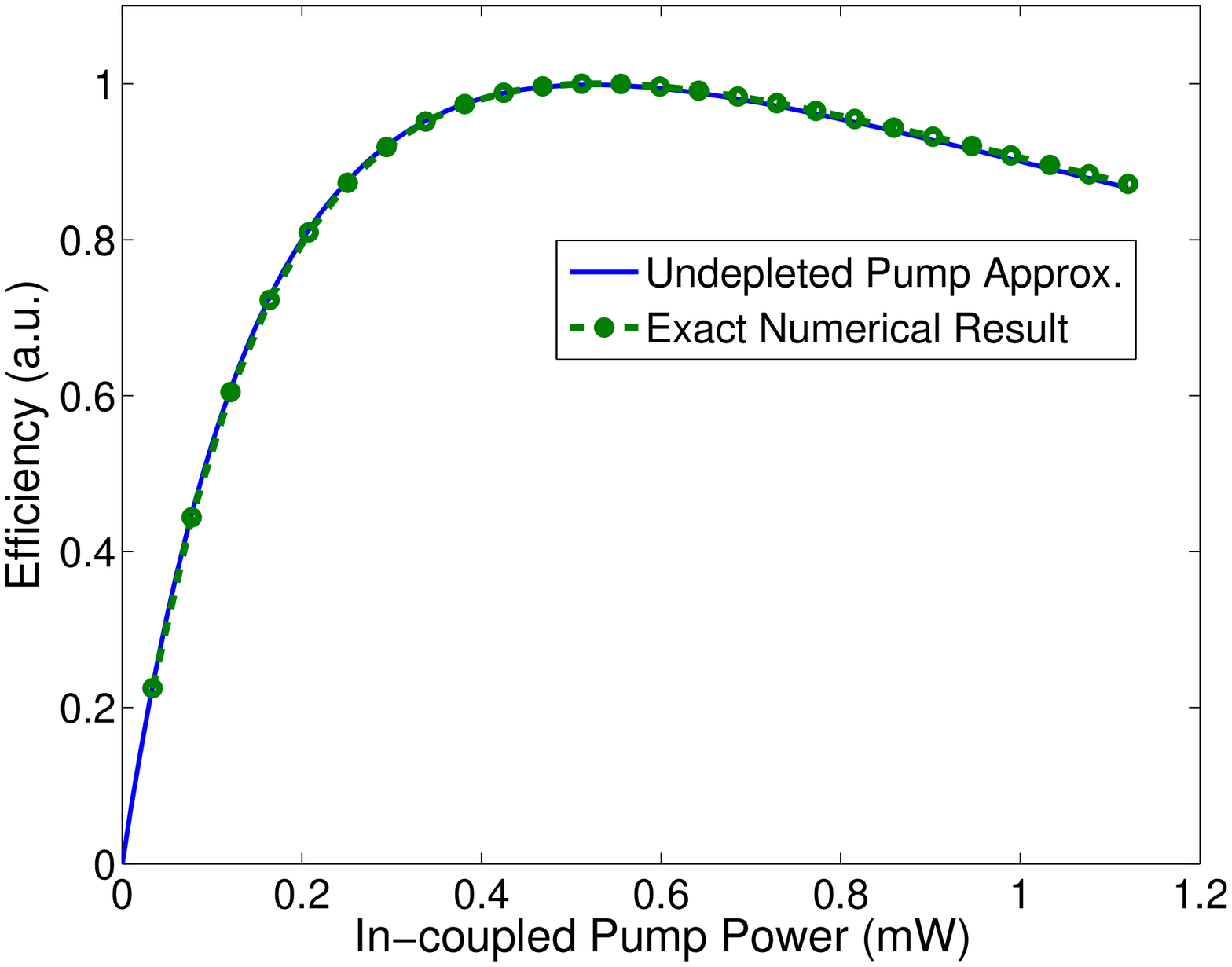}
}
%\vspace*{-0.2in}
\caption[]{\label{fig:compare} The dotted line is the exact numerical solution for 2.5$\mu W$ signal in-coupled power, and the solid line is the undepleted pump approximation. Their good agreement enables us to estimate $\Omega$ and utilize it for higher signal powers, where the approximation is invalid and exact analytical quasi-static expressions do not exist.}
\end{figure}

For low signal powers, the sum-frequency efficiency can be understood in the undepleted pump regimes (Fig.~\ref{fig:compare}). 

Since Eq. (5) in the main text is valid only in the undepleted pump approximation, we can infer $\Omega$ only for datasets where the signal powers are very small. Additionally, we assume that the inherent nonlinearity is so weak that for both pump and signal powers P $\le$ 5$\mu W$, there is no noticeable conversion (which is justified by all of our data sets, wherein the efficiency is nearly zero in this region). Since we use the same WGM triplet for the entire experiment, and $\Omega$ is an overlap measure between the excited WGMs, we operate under the condition that $\Omega$ is the same for arbitrary input powers.  However, what we can acquire from the measured data is $\Omega$ and $\kappa_{f}$, where $\kappa_{f} = \omega_f(\frac{1}{2Q_f^i}+\frac{1}{2Q_f^c})$. The coupling-$Q$ factors, $Q^c_\mu$, are given by \cite{Gorodetsky99} for a TM WGM in a spherical resonator:
\begin{equation}
Q^c_\mu =\frac{\pi}{2} \frac{\sqrt{n_s}(n_s-1)}{\sqrt{n_c^2-n_s^2}}(R k)^{3/2}\exp({2\gamma d})\label{Qc}
\end{equation}
$\mu=s,p$, where $n_s,n_c$ are the indices of refraction in the resonator and prism respectively, $R$ is the resonator radius, $\gamma = \sqrt{k^2(n_s^2-1)}$, and $d$ is the distance between the prism and resonator. We will use (\ref{Qc}) as an order of magnitude estimate for our spheroidal resonator, and multiply it by $n_s^4$ when computing $Q_{f}^{c}$ for the extraordinarily polarized (TE) 520 nm sum-frequency wave, which is due to the different boundary conditions. Since we operate under critical coupling conditions, i.e., $Q_s^{c}=Q_s^{i}$, for the $\lambda_{s}=780$ nm signal wave, we can acquire an empirical value for $d$. In our experiment, for $Q_s^{c} = Q_s^{i} = 4.48\times10^7$ leads to $d = 70.5$ nm. For this distance, for the 520 nm wave, the calculated $Q_{f}^{c} = 5.34\times 10^9$. From data fitting we find  $Q_f^{i}=3.25\times10^7$, which leads to the undercoupling factor $Q_{f}^{c}/Q_{f}^{i}=164$. This estimated intrinsic Q is very close to an earlier reported \cite{furst10} value of $Q^{i} = 4\times10^7$ in the same material at a close wavelength of 532 nm. Let us point out that in lithium niobate a lower Q is indeed expected at a shorter visible wavelength.

Therefore, the internally generated green light was coupled out very inefficiently. This leads us to believe that the photorefractive damage induced by the green light could have been significant. Moreover, from Eq. (1), we see that any changes in the index of refraction due to thermo-optic effects, light-induced charge transport, or even the electro-optic effect due to charge movement on the periphery of the resonator can significantly affect $Q^c_\mu$. The exponential dependence on $d$ places stringent stability criteria, which may have been breached due to the mechanical hysteresis in the brass-oven that was employed to allow for temperature-tuning, leading to a time-varying $Q^c_\mu$. In fact, by studying the transmission curves over the duration of the experiment, we have seen that the total $Q$ of the resonator does change.  Therefore, we conclude that the $\Omega$ we acquire by assuming that $\Delta_f  = 0$ and by also neglecting these concurrent effects,  might be different slightly from the value acquired when these effects are accounted for. This may further explain why the empirically measured internal conversion efficiency is lower by a factor of 50. 

\section{Observing and fitting the WGM spectra}

\begin{figure}[htb]
%\vspace*{-0.1in}
\centerline{ %\hspace*{-0.5in}
\includegraphics[width=9cm]{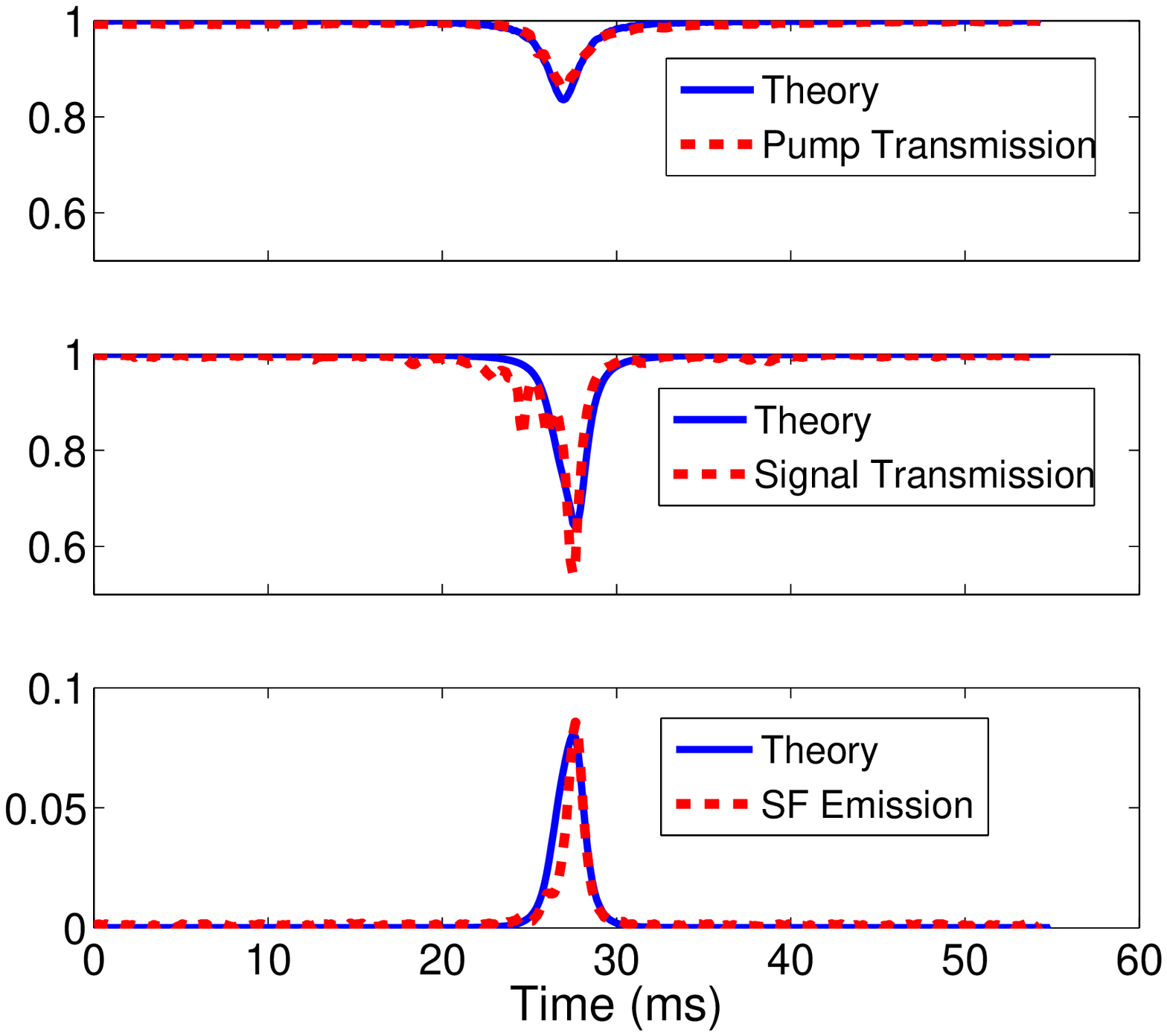}
}
%\vspace*{-0.2in}
\caption[]{\label{fig:traces}Dotted lines are the observed signal, pump and SF spectra, and solid lines are theoretical fits.}
\end{figure}

As discussed in the main part, our measurement produced the signal and pump transmission spectra, and the SF emission spectra, such as shown in Fig.~\ref{fig:traces}.
Having estimated the $\Omega, \kappa_{f}$ parameters from the experimental data, we were able to acquire parameter-free fits for the transmission spectra of the resonator, also shown in  Fig.~\ref{fig:traces}. 

\section{WGMs overlap and SFG channels efficiency}

Spherical symmetry may be a good approximation for determining the eigenfunction shape (but not always the eigenfrequencies) in a spheroid WGM resonator whose aspect ratio is not too large, because the variation of the boundary from a true sphere in the region of significant WGM field is minute.
Due to the spherical symmetry the phase matching in WGM resonators corresponds to conservation of photons' orbital momenta \cite{Kuo11,furst10,furst10pdc,Kozyreff08}. If $L\gg 1$ the overlap integral in $\Omega$ breaks up into a product of the angular and radial part, former depending on $L-m$ and latter on the $q$ for all three modes. Each part also weakly depends on wavelengths and the three values of $L$. The angular parts give rise to the SFG selection rules corresponding to the Clebsch-Gordan coefficients. In particular, they enforce
\begin{eqnarray}
m_p+m_s&=&m_f,\\
L_p+L_s&\geq&L_f,\\
L_p+L_s+L_f&=&2N,
\end{eqnarray}
where $N$ is an integer.
The radial parts lead to no particular selection rules, however as we will see they strongly favor the cases when $q_p+q_s\approx q_f$. 

For mm-size resonators the orbital numbers $L,m$ are large, and evaluating the overlap integrals with Legendre polynomials and Bessel functions of such orders is impractical. Appropriate asymptotic approximations need to be made. Let us introduce 
\begin{equation}
\Psi(r,\theta,\varphi)\approx \Psi_{ang}(\theta)\Psi_{rad}(r)e^{im\varphi}\label{asymp}
\end{equation}
inside the resonator, and $\Psi(r,\theta,\varphi)\equiv 0$ outside, thereby neglecting the evanescent field. The angular asymptotic for (\ref{asymp}) is given by \cite{Louyer2005}
\begin{equation}
\Psi_{ang}(\theta)=\frac{N_a}{\sqrt{2\pi}}H_{L-m}(\sqrt{L}\cos(\theta))\exp\left\{-\frac{1}{2}L\cos(\theta)^2\right\},\label{angular}
\end{equation}
and radial asymptotic is
\begin{equation}
\Psi_{rad}(r)=\frac{N_r}{\sqrt{r}}{\rm Ai}\left(\alpha_q\frac{L-k_r(L+1/2,q)r/R}{L-k_r(L+1/2,q)}\right).\label{radial}
\end{equation}
In (\ref{angular}) $H_{L-m}$ is the Hermite polynomial of the order $L-m$, and $N_a$ is the normalization factor,
\begin{equation}
N_a^{-2}= \int_{-1}^1 H_{L-m}^2(\sqrt{L}x)\exp\left\{-Lx^2\right\}dx.\label{Na}
\end{equation}
In (\ref{radial}), $R$ is the resonator radius, Ai is the Airy function, $\alpha_q$ is its $q$th root (positive value), dimensionless wave number can be approximated \cite{Gorodetsky} as
\begin{equation}
k_r(L,q)\approx L+\alpha_q(L/2)^{1/3},
\end{equation}
and the normalization factor is
\begin{equation}
N_r^{-2}=R^3 \int_{0}^1{\rm Ai}^2\left(\alpha_q\frac{L-k_r(L+1/2,q)x}{L-k_r(L+1/2,q)}\right) xdx.\label{Nr}
\end{equation}
The factors (\ref{Na}) and (\ref{Nr}) provide the asymptotic wavefunction (\ref{asymp}) normalization: $\int dV |\Psi(r,\theta,\varphi)|^2=1$.

% \textbf{Experimental setup!!!}
\begin{figure}[htb]
%\vspace*{-0.1in}
%\centerline{ %\hspace*{-0.5in}
%\input epsf
%\setlength{\epsfxsize}{3.5in} \epsffile{SFG_Setup.eps}
\includegraphics[width=9cm]{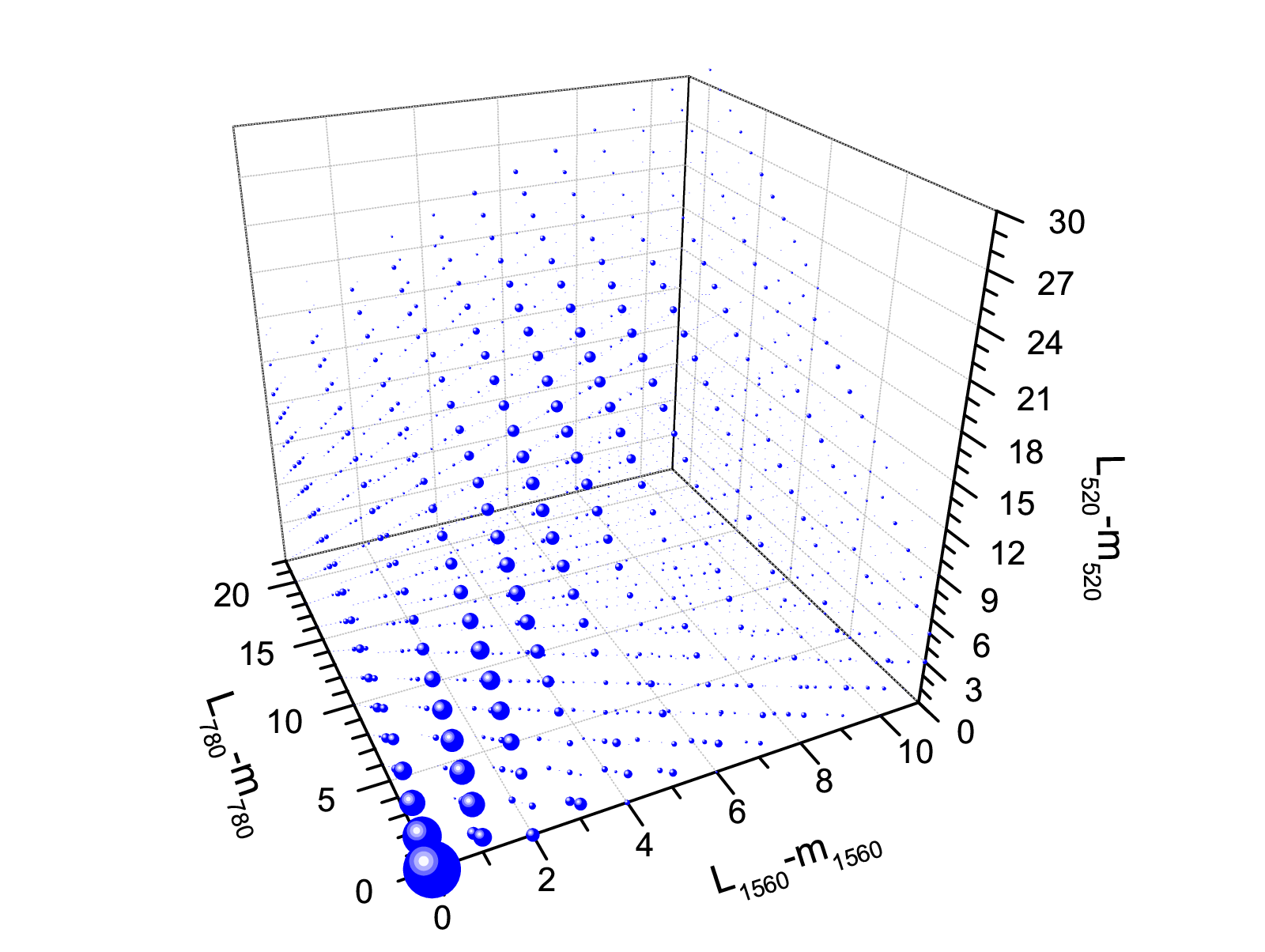}
\includegraphics[width=9cm]{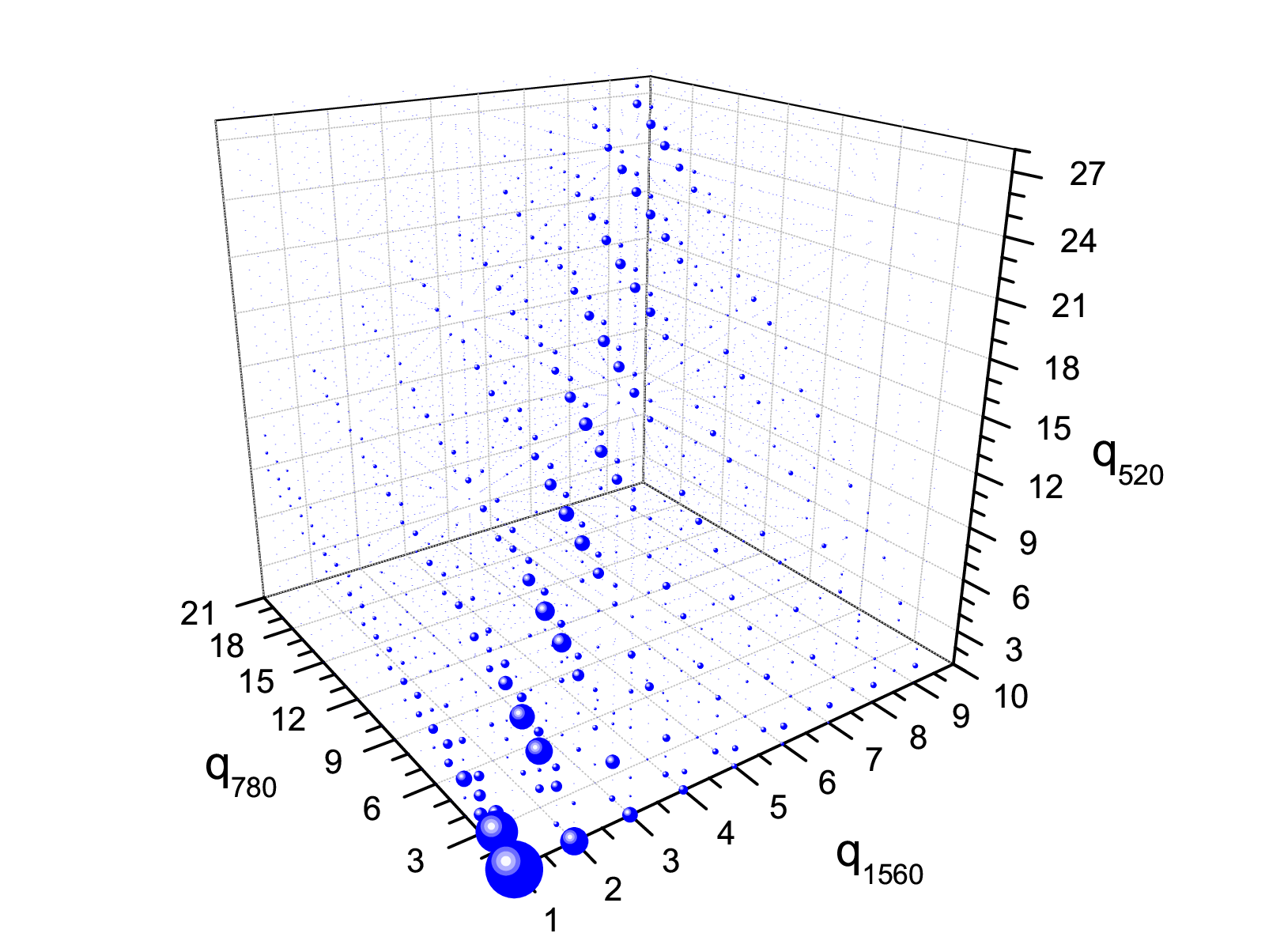}
%}
%\vspace*{2.5in}
\caption[]{\label{fig:ovlp}The absolute-square of the angular (top) and radial (bottom) parts of the WGMs overlap integral are represented by the dot size as a function of the respective mode numbers. In both plots, the largest dot is normalized to unity.}
\end{figure}

Approximation (\ref{asymp}) allows us to evaluate the radial and angular overlap integrals separately. These overlap factors are shown in Fig.~\ref{fig:ovlp} in the normalized form.
To determine the overlap integral for a particular channel one needs to take the appropriate angular part from the top of Fig.~\ref{fig:ovlp} and multiply it by the appropriate radial part from the bottom of Fig.~\ref{fig:ovlp}. For the fundamental modes, i.e. such that $q_p=q_s=q_f=1$, $L_p-m_p=L_s-m_s=L_f-m_f=0$ (the most efficient SFG channel for our wavelengths and resonator size) $|\int dV \Psi_{f}\Psi^{*}_{p}\Psi^{*}_{s}|^2\approx 2.66\times 10^6 \,\rm{cm}^{-3}$. For comparison, the fundamental WGM mode volume in our resonator ranges from approximately $1\times 10^{-7} \,\rm{cm}^{3}$ for the SF to $4\times 10^{-7} \,\rm{cm}^{3}$ for the pump.

\begin{figure}[htb]
%\vspace*{-0.1in}
\centerline{ %\hspace*{-0.5in}
\includegraphics[width=10cm]{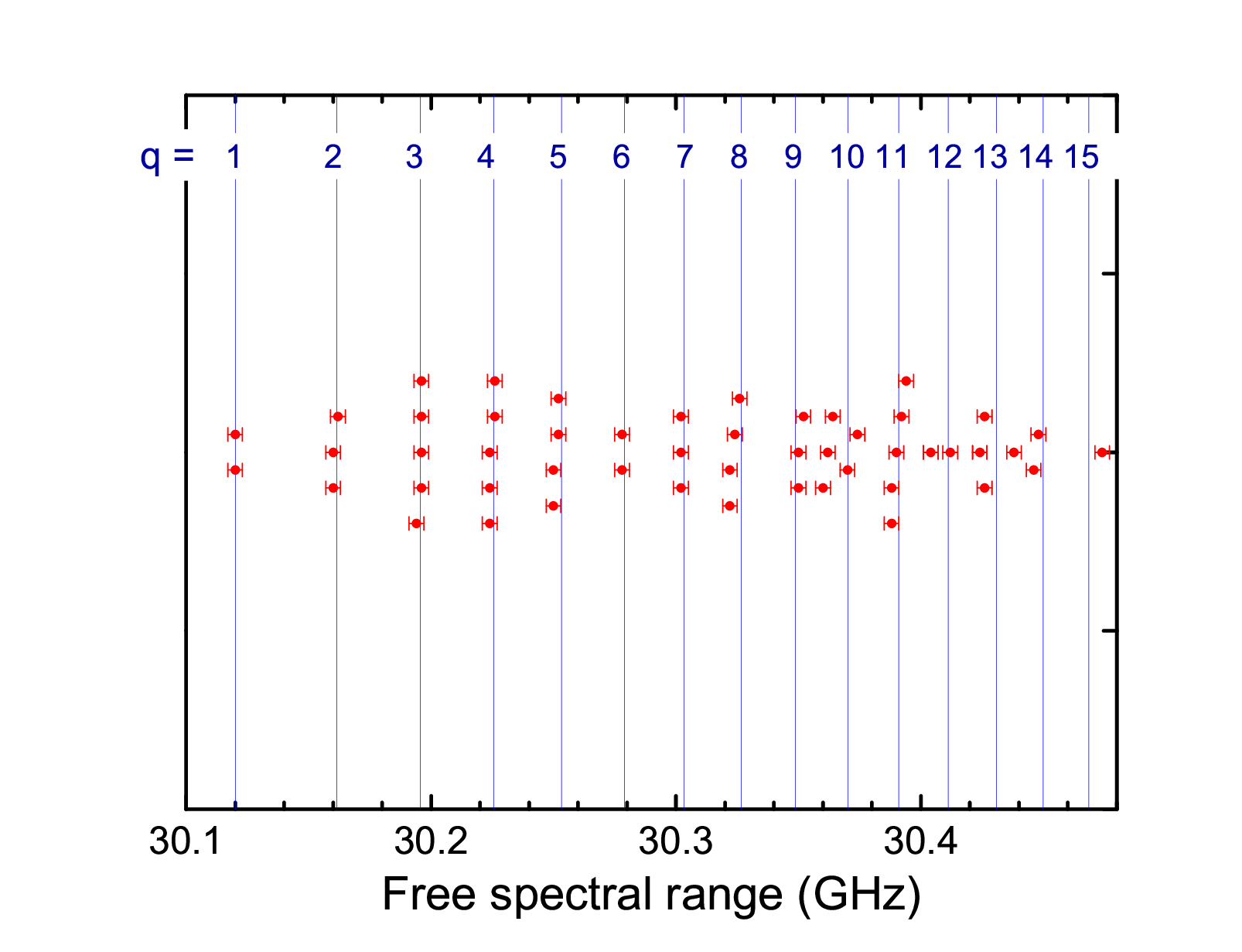}
}
%\vspace*{-0.2in}
\caption[]{\label{fig:FSR}Validation of the $q$-identification approach based on the FSR measurement: vertical lines are theoretical values of FSR for the $q$'s as labeled, data points are the measurement results.}
\end{figure}

This analysis is practically useful only if we can identify the mode numbers for a given mode from a WGM spectrum. As we have discussed in the main part of the paper, identification of the $q$'s is particularly important. Since the technique described therein provided only a limited number of data points at the pump wavelength for our working resonator, and had even more limited success at the signal wavelength, we feel compelled to demonstrate its robustness in a separate measurement with a different test resonator. The test resonator was made from the same material but had a larger radius, $R \approx 700\; \mu$m, and a denser WGM spectrum. The denser spectrum has allowed us to measure the FSR for 50 modes within a single FSR. The results of this measurement carried out with the pump laser are shown in Fig.~\ref{fig:FSR}. Here again, we have forced the $q=1$ theoretical value to match the two lowest-FSR modes by correcting the resonator radius value from $700\; \mu$m to $699.3\; \mu$m. This has achieved a remarkable agreement up to the $q = 9$. At higher $q$'s the WGM coupling contrast is considerably reduced, and the measurements become unreliable. 

\section{Phase matching temperature}

Based on the previous section analysis, we can rank various SFG channels by their efficiency. We then numerically find the dependence of the frequency detuning $\Delta\omega =\omega_f-\omega_s-\omega_p$ on temperature and determine the phase matching ($\Delta\omega =0$) temperature as well as the WGM frequencies $\omega_f,\,\omega_s,\,\omega_p$ at that temperature, for each channel from the ranked list. 

\begin{figure}[htb]
%\vspace*{-0.1in}
\centerline{ %\hspace*{-0.5in}
\includegraphics[width=9cm]{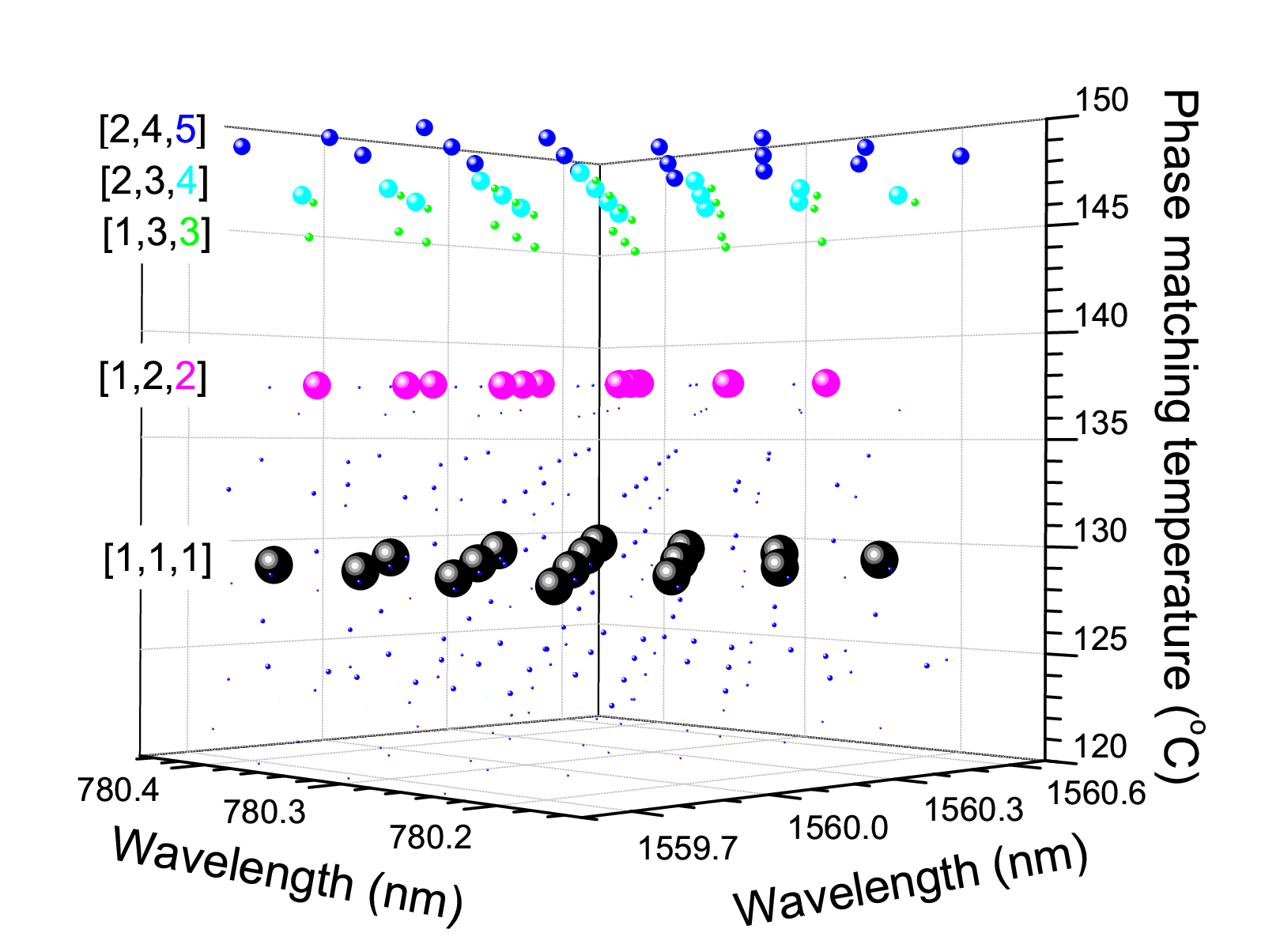}
}
%\vspace*{-0.2in}
\caption[]{\label{fig:layers}Main equatorial SFG channels $[q_s,q_p,q_f]$ form distinct ``layers" in the temperature-wavelength space. 
Within each layer, the channels form the rows of constant $L_s$ and $L_p$. 
Dot sizes represent the channels efficiency.}
\end{figure}

In Fig.~\ref{fig:layers} we show the theoretical prediction for phase matching temperatures in the range between 120 and 150 $^\circ$C. The pump modes in this simulation have been assumed equatorial $(L_p-m_p=L_s-m_s=0)$ and with $q_{s,p}<8$. Both assumptions are justified by the high contrast of the observed WGMs and by the FSR measurements discussed in the main part of the paper. No assumptions have been made about possible SFG WGMs. From Fig.~\ref{fig:layers} we see that the most efficient $[1,1,1]$ SFG channels have the phase matching temperature near 129 $^\circ$C, far from other channels of significant efficiency. However, there are also many ``minor" SFG channels at this temperature. In our experiment, we indeed observed several other SFG channels with the efficiency at least an order of magnitude below than the one we have been using.

% from Numeric simulations\Tuning

\begin{figure}[htb]
%\vspace*{-0.1in}
\centerline{ %\hspace*{-0.5in}
\includegraphics[width=9cm]{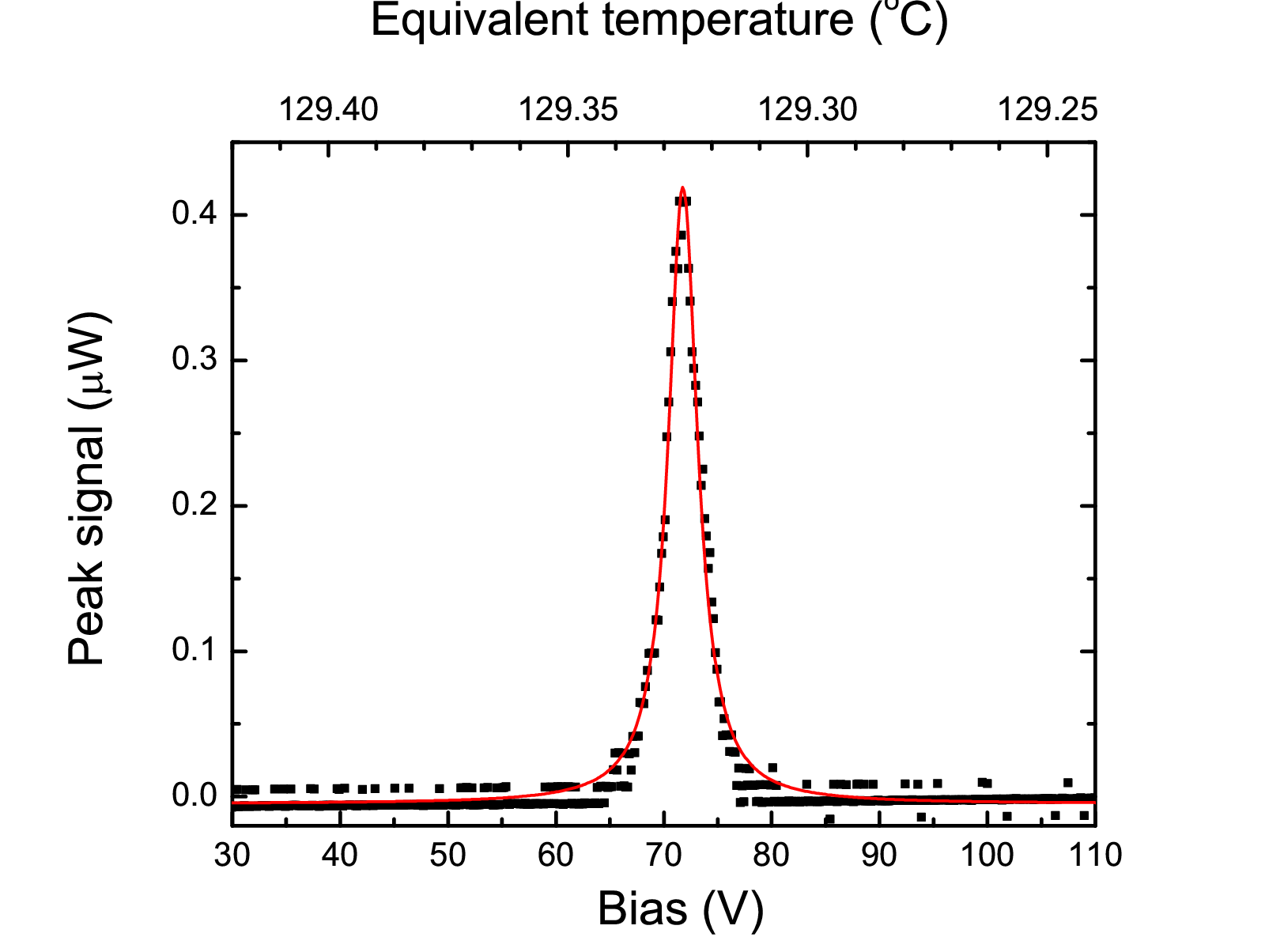}
}
\caption[]{\label{fig:dT}The SFG phase matching temperature width measured by a bias voltage sweep is found from a Lorentzian fit to be approximately 7 milidegrees.}
\end{figure}

The phase matching temperature \emph{width} is determined by the relative drifts of the pump, signal and sum-frequency WGMs with varying temperature. These in turn are determined by the thermal expansion and thermal refractivity of Lithium niobate, and by the mechanical aspects of the resonator mounting. To measure the temperature width of the phase matching, we first calibrated the temperature variation in units of the electro-optical bias voltage variation by a compensation technique, and then recorded the SFG signal vs. the bias voltage sweep. The result of this measurement is shown in Fig.~\ref{fig:dT}. We wee that the phase matching is achieved within a rather narrow temperature range of approximately 7 mK. 

% from \WGM switch\2012 data\Dec 10 Tcal
